\documentclass[final]{cvpr}

\usepackage{times}
\usepackage{epsfig}
\usepackage{graphicx}
\usepackage{amsmath}
\usepackage{amssymb}
\usepackage{subfigure}
\usepackage{overpic}

\usepackage{bm}
\usepackage{algorithm}

\usepackage{algorithmicx}
\usepackage{algpseudocode}

\usepackage{enumitem}
\setenumerate[1]{itemsep=0pt,partopsep=0pt,parsep=\parskip,topsep=5pt}
\setitemize[1]{itemsep=0pt,partopsep=0pt,parsep=\parskip,topsep=5pt}
\setdescription{itemsep=0pt,partopsep=0pt,parsep=\parskip,topsep=5pt}

\usepackage[pagebackref=true,breaklinks=true,colorlinks,bookmarks=false]{hyperref}

\def\cvprPaperID{159} 
\def\confYear{CVPR 2020}

\graphicspath{{figures/}}

\newcommand{\Ignore}[1]{}

\newcommand{\NullFigure}[3]{}

\title{Real-time Rendering and Editing of Scattering Effects for Translucent Objects}

\author{papers 0578}
\author{Rui Wang ~~~~~ Wei Hua ~~~~~ Yuchi Huo ~~~~~ Hujun Bao 
}

\begin{document}

\maketitle

\begin{abstract}
The photorealistic rendering of the transparent effect of translucent objects is a hot research topic in recent years. A real-time photorealistic rendering and material dynamic editing method for the diffuse scattering effect of translucent objects is proposed based on the bidirectional surface scattering reflectance function's (BSSRDF) Dipole approximation. The diffuse scattering material function in the Dipo le approximation is decomposed into the product form of the shape-related function and the translucent material-related function through principal component analysis; using this decomposition representation, under the real-time photorealistic rendering framework of pre-radiative transmission and the scattering transmission to realize real-time editing of translucent object materials under various light sources. In addition, a method for quadratic wavelet compression of precomputed radiative transfer data in the spatial domain is also proposed. Using the correlation of surface points in the spatial distribution position, on the premise of ensuring the rendering quality, the data is greatly compressed and the rendering is efficiently improved. The experimental results show that the method in this paper can generate a highly realistic translucent effect and ensure the real-time rendering speed.

\end{abstract}

Photorealistic rendering of  translucent objects is very important for computer graphics, because many objects in our daily life have translucent properties, such as marble, skin, cheese, bread, snow, etc. Translucency is a special optical effect formed by multiple scattering and partial absorption after light hits the surface of an object. Due to the different degrees of light scattering and absorption, different substances present different translucent visual effects. With the development of computer application requirements, in many applications, such as games, film special effects, industrial design and other fields, there is an increasing demand for quickly generating highly realistic images of translucent objects. However, due to the complexity of the light scattering process, it is very challenging to use a computer to realize the real-time realistic rendering of the translucent effect. In recent years, a series of fruitful researches have been carried out in the domain, ranging from the direct calculation of physics-based subsurface scattering models, to the use of various approximate acceleration algorithms \cite{r1,r2,r3,r4}, and the introduction of pre-computation into the rendering of translucent objects. Real-time rendering of translucent effects \cite{r5,r6}. However, there is still a lack of effective methods for real-time editing of translucent materials of objects, especially in multiple light source environments and variable viewing angles, to achieve real-time realistic rendering of translucent objects with changing materials.

In view of this situation, this paper proposes a method to change the translucent material in real time under a variety of light sources (ambient light, point light source, directional light, area light source, etc.) based on the bidirectional surface scattering reflectance function (bidirectional surface scattering reflectance function). Dipole approximation of surface scatte ring reflectance distribution function, BSSRDF). Under the framework of precomputed radiative transfer, by decomposing the diffuse scattering material function and using the method of quadratic wavelet compression on the radiative transfer data in the spatial domain, the real-time rendering of highly realistic translucent effects and the translucent material are realized. real-time editing.

\section{Related Work}

\textbf{Translucent Objects Rendering.} Translucent effects a form of reflection of light of real world objects, which results from the scattering of light beneath the surface of an object. It can be described by BSSRDF in computer graphics to descibe the light transmission that creates a translucent effect. However, since BSSRDF only describes the relationship between the shooting person and the light emitted from the translucent object, and does not correspond directly to a physical quantity that describes the effect of translucency (such as to light
scatter, absorptivity, etc.). Therefore, for a long period of time, the rendering of translucent objects requires complex calculations to simulate the transmission of light in materials with specific scatter and absorptivity \cite{r2,r3,r4}, and the rendering efficiency is low. Until Jensen et al. \cite{r1} expressed BSSRDF as an analytical form defined by physical parameters such as scattering degree and absorptivity through Dipole approximation, the rendering speed for semi-transparent objects was greatly accelerated \cite{r7}. Even so, the rendering of translucent effects is still far from meeting the needs of real-time applications. On the basis of the Dipole method, Hao et al. \cite{r5} and Wang et al. \cite{r6} speed up the the rendering of translucent effect achieves interactive and real-time rendering respectively. However, these algorithms need to know the translucent material parameters of the object during preprocessing, and cannot change the translucent material in real time when rendering. Xu et al. \cite{r8} further proposed a method to realize real-time editing of semi-transparent materials, but it needs to limit the drawing viewpoint or light source direction. In addition, the method in this paper also has greater advantages in the error control of the calculation and the storage capacity of the pre-calculated data.

\textbf{Precomputed Radiance Transfer}. The precomputed radiance transfer method is a real-time computing framework used to solve the realistic rendering of objects under complex light sources in recent years. It was first proposed by Sloan et al. \cite{r9}. The basic idea of this method is to decompose the function in the rendering equation into a specified set of basis functions, and use the linear combination of these basis functions to approximate the original function. When plotting in real time, take advantage of the good properties of integration between specified basis functions (for example, the orthogonality between spherical harmonic basis functions \cite{r9} and wavelet basis functions \cite{r10} or between certain spherical radial basis functions \cite{r11} that are easy for integrals) to speed up the integral operation of drawing equations. The computational modification transfer method is a hot research topic in recent years: from the beginning, it only supports ambient light, and extends to support local and regional light sources \cite{r12}, BRDF, to gradually support the editing and modification of dynamic BRDF \cite{r13,r14}. The method in this paper takes advantage of the precomputed radiance transport framework and extends it to real-time editing of translucent object materials.

\textbf{Spatial-temporal compression for Precomputed Radiance Transfer}. The predecessors mainly used clustered principal component analysis \cite{r15}, clustered tensor approximation \cite{r11} and other methods to compress the precomputed radiance transmission data. The disadvantages of using principal component analysis methods are that the calculation speed is slow and the stability of numerical calculation is more sensitive. We propose a method to further compress the data by transforming the 3D object surface parameters into a specific parameter domain and performing a wavelet transform on it.
This method has faster calculation speed and better compression ratio.

\section{Translucent material function approximation based on PCA}

\subsection{Approximating BSSRDF via Dipole}

\begin{figure*}
    \centering
    \includegraphics[width=0.8\textwidth]{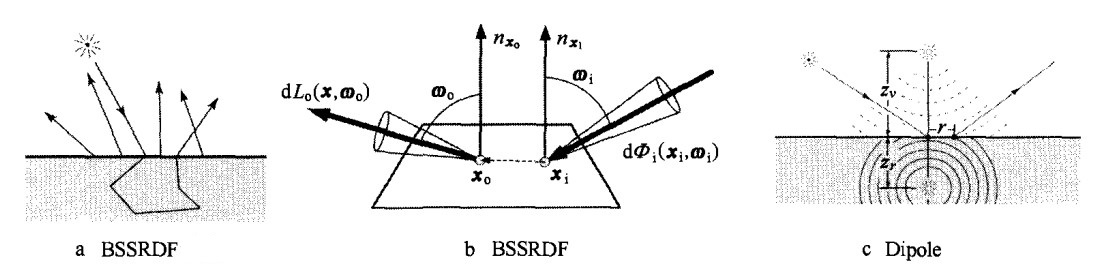}
    \caption{BSSRDF and Dipole Approximation.}
    \label{fig1}
\end{figure*}

According to the Dipole rendering for translucent objects like \cite{r1}, the rendering equation for the scattering effect of a translucent object is
\begin{equation}
    \begin{aligned}
    L_o(x_o,\omega_o)=\int_A \int_\Omega S(x_i,\omega_i;x_o,\omega_o)\cdot\\
    L_i(x_i,\omega_i)(n\cdot \omega_i)d\omega_i d A(x_i),\\
    \end{aligned}
\end{equation}
where $L_o(x_o,\omega_o)$ is the outgoing ray on $X_o$ along $\omega_o$, $S(x_i, \omega_i, X_o, \omega_o)$ is BSSRDF, and $L_i(x_i, \omega_i)$ is the incident light energy of the point $x_i$ along direction $\omega_i$ near $x_o$. $(n\cdot \omega_i)$ reflects the incident angle of the incident light, $\Omega$ is the unit sphere defined by the above light energy and reflective function, and $A$ is the area of the object surface. $S(x_i, \omega_i; x_o, \omega_o)$ describes the relationship between the luminous flux $\Phi_i(x, \omega_i)$ and $L_o(x_o,\omega_o)$ as shown in Figure \ref{fig1} (b). $S(x_i, \omega_i, X_o, \omega_o)$ is usually expressed as the synthesis of single scattering effects and diffuse scattering, respectively, namely
\begin{equation}
    \begin{aligned}
    S(x_i,\omega_i;x_o,\omega_o)=S^{(1)}(x_i,\omega_i;x_o,\omega_o)+\\
    S_d(x_i,\omega_i;x_o,\omega_o).\\
    \end{aligned}
\end{equation}

Generally, the translucent effect of solid matter in nature is mainly caused by diffuse scattering, while liquid translucent matter is mainly manifested as single scattering \cite{r16}. For objects with diffuse scattering and uniform material, Jensen et al. \cite{r1} proposed a Dipole approximation based on diffuse scattering theory, as shown in Figure \ref{fig1} (C). The Dipole approximation assumes that the object has infinite depth and infinite surface. After the ray hits the plane point $x_i$, the diffuse scattering of the ray on the plane can be approximated as he effect of direct illumination of the plane by the two virtual light sources (Dipole points) directly above and below the surface $x_i$. The position of the two virtual light sources relative to the surface of the object is determined by the material function of the semi-transparent object, that is,

\begin{equation}
    \begin{aligned}
    S(x_i,\omega_i;x_o,\omega_o;\sigma)=\\
    \frac{1}{\pi}F_t(\eta,\omega_i)R(\|x_i-x_o\|_2,\sigma)F_t(\eta,\omega_o),\\
    \end{aligned}
\end{equation}

where $F_t$ is the Fresnel transmission coefficient at point $x_i$ and $x_o$; $R_d(\|x_o-x_i, \sigma)$ is the diffuse scattering function, which can be approximated by Dipole as the analytical form related to the material coefficient of the translucent object

\begin{equation}
    R_d(r,\sigma)=\frac{a'}{4\pi}[z_r(\sigma_{tr}+\frac{1}{d_r})\frac{e^{-\sigma_{tr}d_r}}{d^2_r}+z_v(\sigma_{tr}+\frac{1}{d})\frac{e^{-\sigma_{tr}d_v}}{d^2_v}],
    \label{eq1}
\end{equation}
where $r=\|x_o-x_i\|$ is the distance between $x_i$ and $x_o$; $\sigma'_s=(1-g)\sigma_s$ and $\sigma'_t=\sigma_s+\sigma'_s$ is the scattering attenuation and scattering decay coefficients; $\sigma_a$ is the absorption coefficient; $\sigma=(\sigma'_s,\sigma_a)$ is the material coefficient of the scattering functino $R_d(r,\sigma)$; $a'=\frac{\sigma'_s}{\sigma’_t}$ is the decay ratio; $g$ is the average of scattering cosine term; $\sigma_{tr}=\sqrt{3\sigma_s\sigma'_t}$ is the valid decay coefficient; $z_r=\frac{1}{\sigma'_t}$ and $z_v=z_r(1+\frac{4A}{3}$ are the distance between two Dipole to the surface, where $A=\frac{1+F_{dr}}{1-F_{dr}}$, $F_{dr}$ is the approximated Fresnel term, $F_{dr}=\frac{-1.440}{\eta^2}+\frac{0.710}{\eta}+0.668+0.0636\eta$; $d_r=\sqrt{r^2+z_r^2}$ and $d_v=\sqrt{r^2+z_v^2}$ are the distances from $x$ to two Dipole points. The importance of the Dipole approximation is that it gives an analytical representation of the surface scattering effect. Based on this approximation, this paper mainly discusses the real-time photorealistic rendering and real-time material editing for the diffuse scattering effect of translucent materials under the Dipole approximation.

\subsection{Diffuse Scattering Material Function Decomposition and Calculation}

For static scenes, $r$ in is known in advance, while $\sigma$ needs dynamic changes to realize real-time editing of materials. If one simply sample the value space of $\sigma$, it will bring a huge amount of data, which cannot be practically operated. Therefore, we denote $R_d(r, \sigma)$ as the factorial division of $r$ with the $\sigma$, $R_d(r,\sigma)\approx\sum_{k=1}^K S_k(\sigma)b_k(r)$.

Note that $r(x_i)$ is a one-variable function with a value range of $[0,+\infty)$. According to formula \ref{eq1}, $R_d(r,\sigma)$ function has a negative exponential form. When $r$ is large, its value is approximately $0$. In addition, although the value range of $\sigma$ is $(0, + \infty)$, when $\sigma$ becomes large, the assumption of Dipole approximation will no longer be suitable, and BSSRDF can be considered to degenerate into BRDF. This means that the value of the $\sigma$ variable should be limited within a certain area. We can obtain the decomposed form of the diffuse scatter function by performing a principal component analysis on the matrix $M$, that is,
\begin{equation}
    M=USV.
\end{equation}

\begin{figure*}
    \centering
    \includegraphics[width=0.8\textwidth]{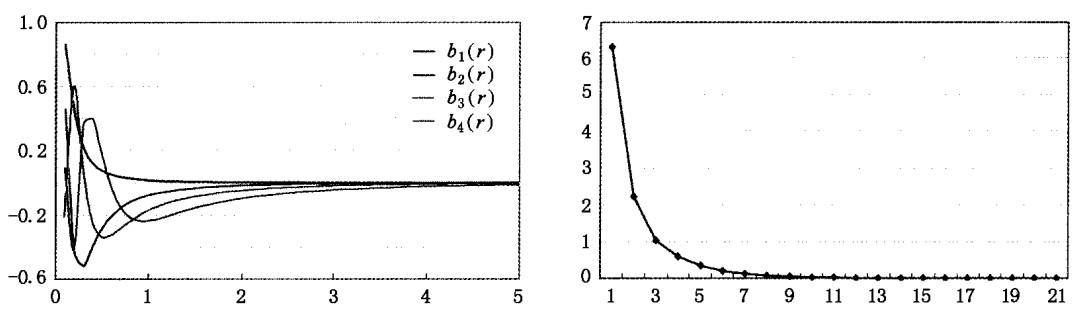}
    \caption{Basis function value and eigenvalue distribution. Left: the basis functions of the first four largest eigenvalues. Right: eigenvalue distribution.}
    \label{fig2}
\end{figure*}

\begin{figure*}
    \centering
    \includegraphics[width=0.6\textwidth]{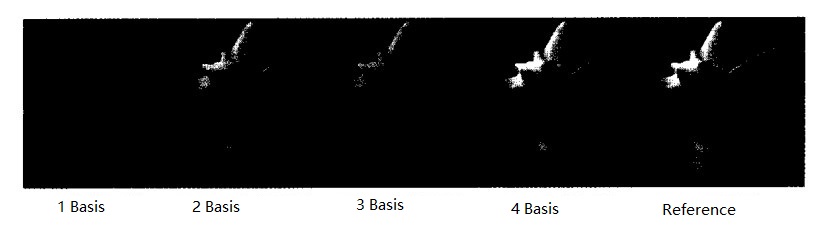}
    \caption{Visualization of approximation using different number basis functions.}
    \label{fig3}
\end{figure*}

According to the nature of pivot analysis, the column vector of $V$ matrix is the discrete representation of basis function $b_k(r)$. Although the $U$ matrix also corresponds to a set of orthogonal functions of $s_k(\sigma)$, in real-time computing, for any given $\sigma$ value, $R_d(r, \sigma)$ can be projected onto the basis function $b_k(r)$ to find the corresponding coefficient $s_k(\sigma)$.

\begin{figure*}
    \centering
    \includegraphics[width=1.0\textwidth]{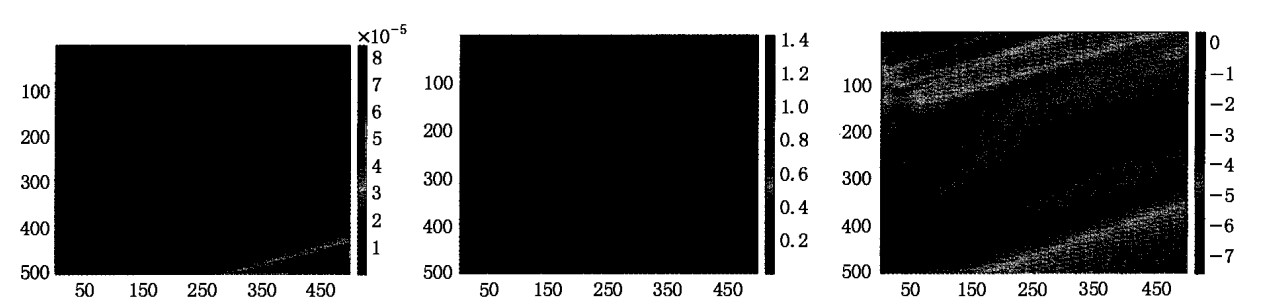}
    \caption{Numerical analysis of approximation error. Left: absolute error under $L^\infty$. Middle: relative error under $L^\infty$. Right: relative error under $\log L^\infty$.}
    \label{fig4}
\end{figure*}

Figure \ref{fig2} shows the value distribution of the first four basis functions $b_1(r)$, $b_2(r)$, $b_3(r)$ and $b_4(r)$ decomposed from the matrix $M$. The right side shows the values of the first 20 eigenvalues arranged from high to low. It can be found that the eigenvalues after No. 10 are very close to 0. In practical applications, the number of basis functions we choose is generally 12. Figure \ref{fig3} shows the visualization results of the translucent object after approximating the diffuse scattering function $R_d(r, \sigma)$ with different numbers of basis functions. It can be seen that when 12 basis functions are used to approximate $R(r, \sigma)$, its Visual errors are hard to detect. It can be seen from Figure \ref{fig4} (a) and \ref{fig4} (b) that the diffuse scattering material function $R(r, \sigma)$ can be well approximated by using 12 basis functions decomposed by PCA.

Compared with Xu et al \cite{r8} of approximating $R(r, \sigma)$ by selecting piece-wise polynomial function, the method in this paper has the following advantages: 1) In the case of the same number of basis functions, the approximation error of the method in this paper is smaller. For example, when there are 15 basis functions, the $L^2$ error of the method in \cite{r8} is 0.005 , while our error is less than 0.0001. 2 ) The basis function selected according to the size of the eigenvalue can easily realize the approximation error
However, the method in \cite{r8} did not give the relationship between the number of basis functions and the approximation error using a piece-wise polynomial approximation. 3) The basis functions obtained by the principal component analysis are orthogonal, and the real-time calculation of the coefficient $s_k(\sigma)$ is faster. However, in the method in the literature \cite{r8}, due to the non-orthogonality of the slice function, a linear system of equations needs to be solved.

\section{Precomputing and Compression of Scatter Transmission}

We substitute the decomposed form of $R(r, \sigma)$ into the translucent object rendering equation and get

\begin{equation}
    \begin{aligned}
    L_o(x_o,\omega_o)=\frac{1}{\pi}F_t(\eta,\omega_o)\int_A\int_{2\pi}F_t(\eta,\omega_i)R_d(r,\sigma)\cdot\\
    L_i(x_i,\omega_i)(n\cdot \omega_i)d\omega_i dA(x_i)\approx\\
    \frac{1}{\pi}F_t(\eta,\omega_o)\sum_k s_k(\sigma)\int_A E(x_i)b_k(r(x_i))dA(x_i)=\\
    \frac{1}{\pi}F_t(\eta,\omega_o)\sum_k s_k(\sigma)\cdot B_k(x_i,r(x_i)).\\
    B_k(x_i,r(x_i))=\int_A E(x_i)b_k(r(x_i))dA(x_i).\\
    E(x_i)=\int_{2\pi}L_i(x_i,\omega_i)(n\cdot\omega_i)d\omega_i.\\
    \label{eq2}
    \end{aligned}
\end{equation}

It can be seen that the calculation of the outgoing ray can be done in 2 steps: 1) Calculate outgoing ray $E(x_i)$ at point $x_i$ on the lower object surface. 2) Calculate $L_o$ according to the outgoing ray $E(x_i)$ of $x_i$. We use the precomputed radiative transfer framework to accelerate the rendering, in which we rewrite Equation \ref{eq2} into matrix form following the method of \cite{r9} as
\begin{equation}
    L_o=\sum_k s_k(\sigma) \cdot (T_k\cdot E) = \sum_k s_k(\sigma)\cdot v_k,
\end{equation}
where $T_k$ is a scattering transportation matrix describes $b_k(\|x_i-x_o\|)$. This matrix describes the scattering and transmission of light by the whole object, the row vector corresponds to the dimension of light transmission, and the column vector corresponds to the spatial dimension of the surface point of the object; $E$ is the outgoing ray vector representing $E(X_i)$; $v_k=T_k\cdot E$.

\subsection{Scattering Transmission Compression}

Following wavelet-based precomputed radiance transfer methods \cite{r10}, we project the outgoing ray vector $E$ and the scattering transfer matrix $T$ to a basis
\begin{equation}
    v_k=(T_k\cdot w^{-1})\cdot(w\cdot E)=T^w_k\cdot ^wE,
    \label{eq3}
\end{equation}
where $w$ is the orthogonal wavelet projection matrix $w^{-1} \cdot w= I$; $T_k^w$ and $^wE$ are the scattering transmission coefficient and outgoing ray coefficient after wavelet projection, respectively. Since the scattered transmission and outgoing rays are both defined on the surface of the object, we use the tree-based hierarchical method proposed by Hasan et al. [\cite{r17} to parameterize the surface points distributed in the 3D space to a regular 2D wavelet domain: $M \times 2^n \times 2^n$ sampling points are evenly distributed on the top of the sample; secondly, according to their distribution in space, the unstructured points on these distributions are divided into $m$ parts, and for each division, the hierarchical method proposed by Hasan is used to construct a balanced quadtree to ensure the uniformity of division; finally, each quadtree is flattened, that is, a regular 2D wavelet domain can be obtained.

\subsection{Spatial Compression}
In addition to the fact that the scatter transmission received at each point on the surface of the object can itself be compressed, the scatter transmission between adjacent points in space is also correlated. Therefore, we can further compress in the airspace. Equation \ref{eq3} can be further expressed as

\begin{equation}
\begin{aligned}
    v_k=w_1^{-1}\cdot(w_1\cdot T_k \cdot w_0^{-1})\cdot (w_0\cdot E)=\\
    w^{-1}\cdot (^{w_1}T_k^{w_0^{-1}}\cdot ^{w_0}E)\\
\end{aligned}
\end{equation}

Since the wavelet transform is linear, we get

\begin{equation}
    L_o = w_1^{-1}\cdot \sum_k s_k(\sigma) \cdot (^{w_1} T_k^{w_0^{-1}} \cdot ^w_0 E)
    \label{eq5}
\end{equation}

where $^{w_1} T_k^{w_0^{-1}}$ indicates that a scattering transmission matrix performs an inverse wavelet transform $w_0^{-1}$ and a wavelet transform $w_1$ at two levels of light transmission and spatial domain, respectively. The wavelet inverse transformation represented by $w_1^{-1}$ in formula \ref{eq5} only needs to be completed once in the last step of rendering, and what is realized is the reconstruction of the surface color of the object.

Since the amount of precomputed data is huge (several gigabytes), we'll do $_o$ first. The transformed wavelet coefficients are subjected to nonlinear compression, and then the compressed vector is subjected to $w_1$ transform. Using 2-step progressive compression can greatly reduce the pre-computation time and preserve the high-frequency effects of the signal as accurately as possible.

An important significance of airspace compression is that after using spatial compression, the included scattering effect does not change greatly with the change of the number of object grid points, and has roughly the same entropy value. Spatial compression can be regarded as the entropy value approximated by compression, ignoring the subtle changes in the surface scattering signal caused by the increase of surface mesh points.

In conclusion, our pre-calculation process includes two steps:

Step 1. Traverse every vertex of the object and every basis that approximates $R_d(r, \sigma)$ function term $b_k(r)$, computes the scattering transfer matrix, and will perform a wavelet transformation, keeping the $n$ item with the largest coefficient stored.

Step2. Perform airspace compression. for each basis function for all vertices in $b_k(r)$ wavelet transforms and compresses the wavelet coefficients retained on the number of terms. In compression, a threshold energy is pre-determined, and certain wavelet coefficients are reserved so that the reserved energy exceeds the threshold.

\section{Real-time Rendering of Translucent Effects}

Since the method in this paper does not depend on the form of direct lighting calculation $E$, it can integrate all the lighting forms that meet the real-time calculation into the above-mentioned framework, such as point light source, directional light, local light source, ambient light and so on. We implement the above models separately and further accelerate rendering computations for some lighting models.

We use the standard Shadow Mapping algorithm to deal with the situation of point light source and directional light, organize the position and normal direction of the surface sampling point as a texture, and calculate the radiation transfer value affected by the shadow in the shader of the graphics card; Then read the data from the graphics card GPU back to the memory, and project it onto the wavelet basis to get $^{w_0}E$.

Using standard wavelet-based precomputed radiance for ambient light \cite{r10}, the transmission method calculates outgoing rays at each sampling point as 
\begin{equation}
    E(x_i)=\int_{2\pi}L_i(\omega_i)(V(x_i,\omega_i)F_i(\omega_i)(n\cdot \omega_i))d\omega_i.
\end{equation}
Express this equation as a matrix, and take the ambient light $L_i(w_i)$ after wavelet transform, we can get
\begin{equation}
    E=V\cdot L_i=(V\cdot w_2^{-1})\cdot (w_2\cdot L_i)=V^{w_2^{-1}} \cdot ^{w_2}L_i,
\end{equation}
where $V$ is the visibility vector at each sample point matrix , which can be obtained by precomputing. Due to the scattering transmission compression, the sample points of the surface have been organized in 2D domain, and after substituting equation it into equation \ref{eq3}, we can further do wavelet transform as
\begin{equation}
    ^{w_0}E = w_o\cdot (V\cdot w_2^{-1})\cdot (w_2\cdot L_i)=^{w_o} V^{w_2^{-1}} \cdot ^{w_2}L_i,
\end{equation}

then we get 
\begin{equation}
    \begin{aligned}
        L_o=w_1^{-1}\cdot \sum_k s_k(\sigma)\cdot (^{w_1}T_k^{w_o^{-1}}\cdot ^{w_o}E)=\\
        w_1^{-1}\cdot \sum_k s_k(\sigma)\cdot (^{w_1}T_k^{w_o^{-1}}\cdot ^{w_o}V^{w_2^{-1}}\cdot ^{w_2}L_i)=\\
        w_1^{-1}\cdot \sum_k s_k(\sigma)\cdot (^{w_1}T_k^{'w_2^{-1}}\cdot ^{w_2}L_i),\\
    \end{aligned}
\end{equation}
where $^{w_1}T_k^{'w_2^{-1}}=^{w_1}T_k^{w_o^{-1}}\cdot ^{w_o}V^{w_2^{-1}}$.

Therefore, we can integrate the pre-computed visibility information into the spatial compression. After pre-computing the radiation transfer matrix $^{w_1}T_k^{'w_2^{-1}}$ including the visibility information and the scattering transmission information, spatially compressing can reduce the storage cost of pre-computed visibility and further improve the speed of real-time rendering.

In order to be able to deal with the situation of the local light source, we introduce the idea of shadow field proposed by Zhou et al \cite{r12}. By calculating the radiation divergence field of the light source, the situation of the local light source is integrated into the processing of the ambient light, so as to realize the lower half of the local light source. Real-time rendering of transparent objects.

The steps of the rendering algorithm are as follows:

Step 1. Calculate the direct illumination $E$ and project it onto the wavelet base to get
$^{w_o}E$.

Step 2. For each $k$ calculate the scattering transfer matrix $^{w_1}T_k^{'w_2^{-1}}$, and its product with $^{w_o}E$, $ ^{w_1}V_k = ^{w_1}T_k^{'w_2^{-1}} \cdot ^{w_o}E$ .

Step 3. Real-time projection of diffuse scattering material functions onto decomposed basis functions, calculate the material parameter $s_k(\sigma)$.

Step 4. Calculate the sum of all $^{w_1}V_k$ with $s_k(\sigma)$ as the weight.

Step 5. Do the inverse transform $w_1^{-1}$ of $w_1$ to get the object surface outgoing ray $L_o$ of the grid point.

It can be seen that if the user needs to adjust the scattering material, he only needs to update Step 3 and Step 4, and the recalculation of the material of the translucent object can be completed quickly to meet the requirements of real-time calculation.

\section{Results}

\subsection{Compression of Scatter Transmission}

\begin{figure*}
    \centering
    \includegraphics[width=1.0\textwidth]{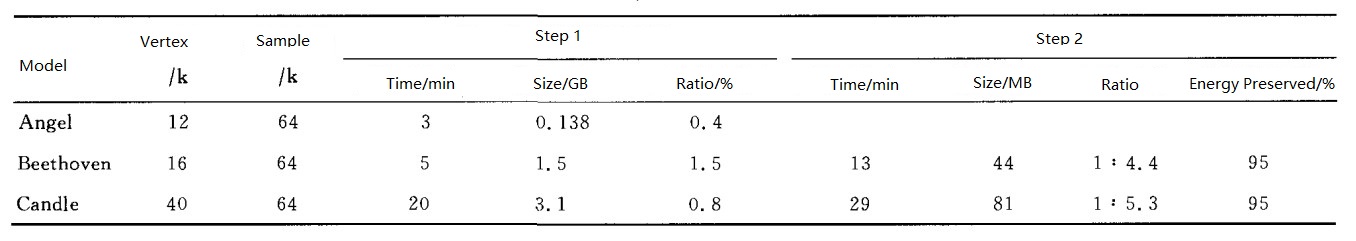}
    \caption{Pre-computation and compression results.}
    \label{tab1}
\end{figure*}

In the compression of the first step, we use the Haar wavelet basis; since the Haar wavelet will produce obvious drawing errors when compressing the spatial domain, in the compression of the second step, a smoother and higher-order wavelet is selected. . We find that the Daub echies 9/7 wavelet can meet the requirements well, and even at a very high compression rate, it will not produce too much rendering error.

\begin{figure}
    \centering
    \includegraphics[width=0.48\textwidth]{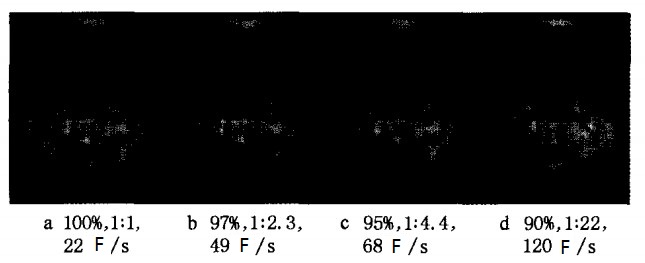}
    \caption{Spatial compression comparison.}
    \label{fig5}
\end{figure}

Figure \ref{tab1} shows the precomputed data and compression results for several different models. The uncompressed original data is very huge, the amount of data is 35 GB ~ 1.1 TB. After the first step of compression, retaining the wavelet coefficient of about $1\%$ can already preserve the effect of light scattering with high quality. The second step is to perform spatial compression: due to the small amount of data in the Angel model, we did not perform spatial compression on it; through experiments on Beeth oven and Candle models, we found that the compression ratio of about 1:4 is preserved, and the The energy of $95\%-98\%$ has little effect on the rendering quality; and the compressed data can be put into the memory, which meets the needs of real-time rendering. Figure \ref{fig5} shows the visual effects under different spatial compression ratios. The numbers in the title of the sub-plots represent: the ratio of energy saving, the compression ratio, and the frame rate of drawing (frame/s). It can be seen that with the increase of the compression ratio and the decrease of the retained energy, the control error will become more and more obvious; and as more and more high-frequency components are discarded, the rendering results become blurred. Such as the shadows on the neck and cheeks of the Beethoven model.

\subsection{Real-time Photorealistic Rendering and Editing of Translucent Materials}

\begin{figure}
    \centering
    \includegraphics[width=0.48\textwidth]{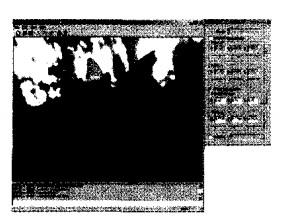}
    \caption{Real-time photorealistic rendering and editing system for translucent materials.}
    \label{fig6}
\end{figure}

\begin{figure}
    \centering
    \includegraphics[width=0.48\textwidth]{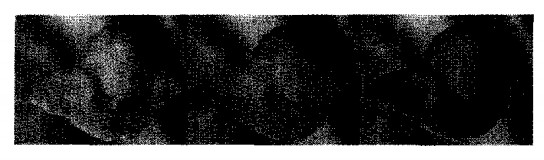}
    \caption{Angel model in ambient light.}
    \label{fig7}
\end{figure}

\begin{figure}
    \centering
    \includegraphics[width=0.48\textwidth]{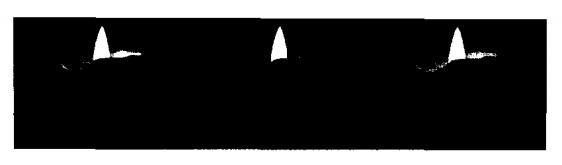}
    \caption{The candle translucent material under local lighting.}
    \label{fig8}
\end{figure}

\begin{figure}
    \centering
    \includegraphics[width=0.48\textwidth]{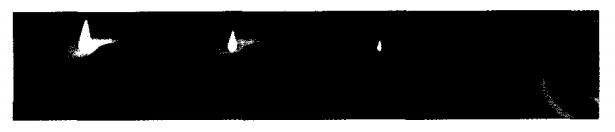}
    \caption{Translucent rendering effect of candles under dynamic light source.}
    \label{fig9}
\end{figure}

\begin{figure}
    \centering
    \includegraphics[width=0.48\textwidth]{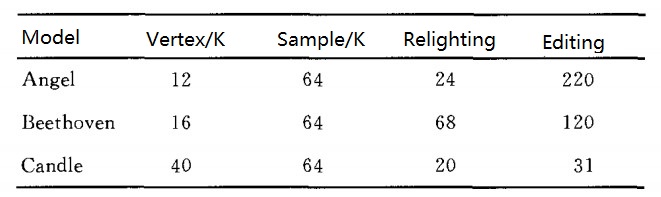}
    \caption{Rendering frame rate.}
    \label{tab2}
\end{figure}

We have developed a system that can easily realize real-time photorealistic rendering and editing of different mesh models, translucent materials, and materials under multiple light source environments, as shown in Figure \ref{fig6}. Figure \ref{fig7} shows the rendering result of the material change of the An gel model under ambient light. Figure \ref{fig8} shows the result of material changes for a candle under a local light source (center flame) (drawing of semi-transparent objects is limited to candles, and the center flame is only a representation). Figure \ref{fig9} shows the translucent effect of the candle model when the light source changes. It can be seen from these results that the method in this paper can obtain high-quality semi-transparent rendering effects with a high degree of realism. Figure \ref{tab2} shows the drawing frame rate of the method in this paper. It can be seen that the method in this paper can achieve real-time speed whether it is the rendering of semi-transparent objects or the editing of transparent material coefficients.

\section{Conclusion}

This paper discusses the technology of real-time photorealistic rendering and material editing based on Dipole approximation for diffuse scattering effect of semi-transparent objects. Firstly, a decomposition approximation method of diffuse scattering material function based on principal element analysis is proposed, and based on this method, the pre-computing and compression methods of scattering transmission are further studied, and the high-fidelity rendering and real-time material editing of semi-transparent objects are realized. The innovation of this method lies in: 1) This paper proposes a method to obtain the optimal approximation of diffuse scattering function through principal element analysis. Compared with other forms of basis functions, the basis functions decomposed by principal component analysis have better properties. Under the same approximation error requirements, fewer basis functions are required and the drawing efficiency is higher. 2) A wavelet-based spatial domain compression method is proposed. Different from the traditional clustering principal component analysis method, the wavelet-based compression method can better preserve the high-frequency information of the signal, and the calculation is simple while maintaining high-quality rendering. At the same time, it greatly reduces the amount of data and improves the drawing speed. It is believed that the method in this paper has application value in the fields of industrial design and the generation of high-fidelity images.

There are some extended articles about applying physical lighting computation in various applications:

\begin{enumerate}
    \item Deep Learning-Based Monte Carlo Noise Reduction By training a neural network denoiser through offline learning, it can filter noisy Monte Carlo rendering results into high-quality smooth output, greatly improving physics-based Availability of rendering techniques \cite{huo2021survey}, common research includes predicting a filtering kernel based on g-buffer \cite{bako2017kernel}, using GAN to generate more realistic filtering results \cite{xu2019adversarial}, and analyzing path space features Perform manifold contrastive learning to enhance the rendering effect of reflections \cite{cho2021weakly}, use weight sharing to quickly predict the rendering kernel to speed up reconstruction \cite{fan2021real}, filter and reconstruct high-dimensional incident radiation fields for unbiased reconstruction rendering guide \cite{huo2020adaptive}, etc.
    \item The multi-light rendering framework is an important rendering framework outside the path tracing algorithm. Its basic idea is to simplify the simulation of the complete light path illumination transmission after multiple refraction and reflection to calculate the direct illumination from many virtual light sources, and provide a unified Mathematical framework to speed up this operation \cite{dachsbacher2014scalable}, including how to efficiently process virtual point lights and geometric data in external memory \cite{wang2013gpu}, how to efficiently integrate virtual point lights using sparse matrices and compressed sensing \cite{huo2015matrix}, and how to handle virtual line light data in translucent media \cite{huo2016adaptive}, use spherical Gaussian virtual point lights to approximate indirect reflections on glossy surfaces \cite{huo2020spherical}, and more.
    \item Automatic optimization of rendering pipelines Apply high-quality rendering technology to real-time rendering applications by optimizing rendering pipelines. The research contents include automatic optimization based on quality and speed \cite{wang2014automatic}, automatic optimization for energy saving \cite{ wang2016real,zhang2021powernet}, LOD optimization for terrain data \cite{li2021multi}, automatic optimization and fitting of pipeline rendering signals \cite{li2020automatic}, anti-aliasing \cite{zhong2022morphological}, etc.
    \item Using physically-based process to guide the generation of data for single image reflection removal \cite{kim2020single}; propagating local image features in a hypergraph for image retreival \cite{an2021hypergraph}; managing 3D assets in a block chain-based distributed system \cite{park2021meshchain}.
\end{enumerate}

\bibliographystyle{ieee}
\bibliography{srbib}

\end{document}